\documentclass[12pt]{article}

\usepackage{amssymb}
\newcommand{\beq}{\begin{equation}}
\newcommand{\eeq}{\end{equation}}
\newcommand{\beqn}{\begin{eqnarray}}
\newcommand{\eeqn}{\end{eqnarray}}

\newcommand{\Tr}{{\rm Tr}}

\newcommand{\be}{\begin{equation}}
\newcommand{\ee}{\end{equation}}
\newcommand{\ba}{\begin{eqnarray}}
\newcommand{\ea}{\end{eqnarray}}
\newcommand{\bdm}{\begin{displaymath}}
\newcommand{\edm}{\end{displaymath}}

\newcommand{\ie}{{\it i.e.\ }}
\newcommand{\eg}{{\it e.g.\ }}

\DeclareMathAlphabet{\mathpzc}{OT1}{pzc}{m}{it}
\def\bea{\begin{eqnarray}}
\def\eea{\end{eqnarray}}
\def\beas{\begin{eqnarray*}}
\def\eeas{\end{eqnarray*}}
\def\sla{\raise.15ex\hbox{$/$}\kern-.57em}

\def\bea{\begin{eqnarray}}
\def\eea{\end{eqnarray}}
\def\de{\partial}

\def\sla{\raise.15ex\hbox{$/$}\kern-.57em}
\def\ie{{\it i.e.}~}
\def\eg{{\it e.g.}~}
\def\ap{{\alpha^\prime}}

\def\cE{{\cal E}}

\def\cI{{\cal I}}

\def\cN{{\cal N}}

\def\cP{{\cal P}}
\def\cQ{{\cal Q}}

\def\cZ{{\cal Z}}

\def\N{$\cal N$}
\begin{document}
\begin{titlepage}
\begin{flushright}
{ROM2F/2009/25}\\
CERN-PH-TH/2009/234 \\
SU-ITP/2009/53
\end{flushright}
\vskip 1cm
\begin{center}
{\Large\bf  Observations on Arithmetic Invariants
and U-Duality Orbits in $\cN =8$ Supergravity}\\
\end{center}
\vskip 2cm
\begin{center}
%%%%%%%%%%%%%%%%%%%%%%%%%  AUTORI  %%%%%%%%%%%%%%%%%%%%%%%%%%%%%%%%%%%%%%%%%
{\bf Massimo Bianchi$^{1,2}$}, {\bf Sergio Ferrara$^{1,3,4}$},
and {\bf Renata Kallosh$^5$} \\
$^1${\sl Physics Department, Theory Unit, CERN \\ CH 1211, Geneva
23,
Switzerland }\\
$^2${\sl Dipartimento di Fisica, Universit\'a di Roma ``Tor Vergata''\\
 I.N.F.N. Sezione di Roma ``Tor Vergata''\\
Via della Ricerca Scientifica, 00133 Roma, Italy}\\
$^3${\sl INFN - Laboratori Nazionali di Frascati \\ Via Enrico
Fermi 40, 00044 Frascati, Italy}\\
$^4${\sl Department of Physics and Astronomy \\
University of California, Los Angeles, CA USA}\\
$^5${\sl Department of Physics, Stanford University, Stanford, CA
94305}
\end{center}
\vskip 1.0cm

%\selectlanguage{american}
\begin{abstract}

We establish a relation between time-like, light-like  and
space-like orbits of the non-compact $E_{7(7)}(\mathbb{R})$
symmetry and discrete $E_{7(7)}(\mathbb{Z})$  invariants.  We
discuss the U-duality invariant formula for the degeneracy of
states $d(\cQ)$  which in the approximation of large occupation
numbers reproduces the Bekenstein-Hawking entropy formula for
regular black holes with $AdS_2$ horizon. We explain why the
states belonging to light-like orbits, corresponding to classical
solutions of ``light black holes with null singularity'', decouple
from the corresponding index.  We also present a separate
U-duality invariant formula for the class of light-like orbits
specified by discrete $E_{7(7)}(\mathbb{Z})$ invariants. We
conclude that the present study of the non-perturbative sector of
the theory does not reveal any contradiction with the conjectured
all-loop perturbative finiteness of D=4 \N=8 supergravity.

\end{abstract}
\end{titlepage}

\tableofcontents

\section{ Introduction}

We analyze the non-perturbative completion of D=4, \N=8 supergravity following the proposal in \cite{Bianchi:2009wj} with the
purpose of understanding the possible implications on the
conjectured UV finiteness of the perturbative theory
\cite{Bern:2009kd, Kallosh:2009db}, based on its relation to \N=4
SYM theory \cite{Bern:1998ug, Bianchi:2008pu}. The important
conclusion from the analysis in \cite{Bianchi:2009wj} was that the
string theory states \footnote{In
\cite{Green:2007zzb}   the argument of the non-decoupling
of string theory states from $D=4$ $\cN =8$ supergravity was based mostly on
Kaluza-Klein monopoles which are 1/2
BPS states.   Other non-perturbative excitations were identified in
\cite{Green:2007zzb}, which preserve less supersymmetry but they are
believed to be less crucial for the non-decoupling argument. } which
do not decouple in $D=4$ \cite{Green:2007zzb}, when viewed as
solutions of classical $\cN =8$ supergravity, expose null
singularity \footnote{It has been suggested in
\cite{Schnitzer:2007rn}
 that the tower of states in \cite{Green:2007zzb} may
 be seen in D=4 via the infinite sum of non-planar Feynman graphs which
 may have  an infinite number of Regge-cuts.}. These states in
non-perturbative $D=4$ $\cN =8$ supergravity all have no mass gap,
they are light and  may become massless at the boundary of the
moduli space $E_{7(7)}/SU(8)$, which lies at infinite distance
from any interior point and where perturbative $D=4$ $\cN =8$
supergravity fails to be valid. The area of the horizon of such
singular solutions with 1/2, 1/4, 1/8 unbroken supersymmetry is
zero due to the vanishing of the quartic Cartan invariant of the
$E_{7(7)}$ symmetry, \ie $\cI_4(\cQ)=0$.

The studies in \cite{Bianchi:2009wj} raised the following issue:
is it possible that the states discussed in \cite{Green:2007zzb}
may be consistently excluded from the four-dimensional theory and
therefore do not affect the UV properties of $\cN =8$ $D=4$
supergravity  or they can be proven to be necessary in $D=4$ and
therefore will affect the perturbative theory? To distinguish
between these two possibilities, it is useful to study the $\cN
=8$ BPS partition function. One should try to understand if the
U-duality symmetry would be broken if the states with null
$E_{7(7)}$ quartic invariant and thus vanishing area of the
horizon were left out. If this were the case, it would suggest
that the conjectured UV finiteness of perturbative theory may be
disproved. On the other hand, if one could show that there are
independent U-duality invariant partition functions for  $\cI_4=0$
states and for $\cI_4\neq 0$ states, this might be viewed  as
further evidence for the conjectured finiteness of $D=4$ $\cN =8$
supergravity. Since such states form separate orbits in the
non-compact $E_{7(7)}$ symmetry one would expect that they should
not mix with each other. Still, in order to support this
expectation, \ie the absence of U-duality anomalies, one should
explicitly find formulae for the degeneracy of states with
$\cI_4=0$ and with $\cI_4\neq 0$ and show that they are separately
U-duality invariant.

There is currently no agreement on whether the analysis of the
$E_{7(7)}(\mathbb{Z})$ symmetry of the partition function or
appropriately modified Witten indices may really serve as an
efficient discriminator between finite and non-finite $D=4$  $\cN
=8$ supergravity. However, since higher loop computations are not
likely to help to find the difference, it seems that careful
derivation of explicit formulae for the degeneracy of $\cN =8$
states and supersymmetric indices is one  way to make progress and
eventually, reach some definite conclusion.

The purpose of this paper is therefore to establish the properties
of $E_{7(7)}(\mathbb{Z})$ invariant partition functions and
appropriately modified Witten indices  and explore the role of the
$\cI_4=0$ states in the computation of the degeneracy of states.
If we knew how to derive the manifestly invariant
$E_{7(7)}(\mathbb{Z})$ partition functions, we could realize the
full microscopic physics of M-theory. It is not surprising,
therefore, that no such explicit formulae are immediately
available. However, certain answers, based on the counting of
states in string theory \cite{Maldacena:1999bp},
\cite{Shih:2005qf}, \cite{Pioline:2005vi} where U-duality is first
broken to some of its subgroups, and afterwards restored, are
available, and we will analyze them.  Our analysis will be  based
on the recent studies in \cite{Sen:2009gy,Sen:2008sp,Sen:2008ta}
where the discrete $E_{7(7)}(\mathbb{Z})$ invariants play a
prominent role.

The importance of supersymmetric indices is due to their
moduli-independence that allows for a robust counting of
particular BPS states. In the simplest cases, Witten indices count
the number of bosonic states minus the number of fermionic states
with assigned charges $\cQ$. The index is defined separately for
supermultiplets preserveing specific fractions of the original
supersymmetry. For instance, the index $B_{14}$  is computed in
\cite{Sen:2008ta} for 1/8 BPS states of $\cN =8$ supergravity as
the difference between the number of bosonic states and fermionic
states of the minimal short multiplet with 28 fermionic zero
modes. States with 1/4 unbroken supersymmetry would contribute to
a different index, $B_{12}$, since these states have 24 fermionic
zero modes and form shorter multiplet. Finally, 1/2 BPS states
contribute to a separate index, $B_{8}$, as they break 16
superymmetries and form an ultrashort multiplet. Therefore the
properties of the indices suggest that the $\cI_4=0$ states which
are 1/4 or 1/2 BPS are decoupled from the degeneracy formula where
1/8 BPS states contribute. We will also see how this decoupling is
realized when the states have particular values of the arithmetic
discrete $E_{7(7)}(\mathbb{Z})$ invariants.

The index $B_{14}$ may include 1/8 BPS states with $\cI_4=0$ as
well as states with $\cI_4\neq 0$, as will be clear from the
properties of the discrete $E_{7(7)}(\mathbb{Z})$ invariants. We
will find out, however, that $\cI_4=0$ 1/8 BPS states form a
separate U-duality invariant orbit whose degeneracy formula will
be given explicitly below.

The  paper is organized as follows. In sec. 2 we describe how to
derive the \N=8 black hole entropy from microscopic state
counting.  The characterization of orbits in terms of discrete
invariants of $E_{7(7)}(\mathbb{Z})$ is addressed in sect. 3. In
sect. 4 we review Sen's analysis
\cite{Sen:2009gy,Sen:2008sp,Sen:2008ta} for 1/8 BPS states and
show that the formulae for  $\cI_4 \neq 0$ states and for  states
with $\cI_4 = 0$ are separately U-duality invariant.  Degeneracy
of 1/2 and 1/4 BPS states admitting a perturbative string
description are considered in sect. 5. In sect. 6, higher
dimensional resolution of the null singularity of `light black
hole' is presented based on the decomposition of the quartic
invariant of D=4 symmetry,  $E_{7(7)}(\mathbb{Z})$,  in terms of
the cubic invariant of D=5 symmetry, $E_{6(6)}(\mathbb{Z})$,  and
in terms of the quadratic invariant of the D=6 U-duality symmetry
$O(5, 5)(\mathbb{Z})$. In sect. 7 we compare the relation between perturbative and non-perturbative sectors of \N=8 supergravity with that on \N=4 Yang Mills theory.
Sect. 8 contains our conclusions and a
discussion of our results.

In appendix A we provide some details on the elliptic functions,
useful for understanding formulae for the degeneracy of states. In
Appendix B we provide the relation between the 28 electric and 28
magnetic charges of D=4 \N=8 supergravity and states of M-theory
compactified on $T^7$ and 27 electric and 27 magnetic charges of
the D=5 \N=8 supergravity and states of M-theory compactified on
$T^6$.

\section{From Degeneracy of States to Bekenstein-Hawking Entropy}

We start by reviewing the derivation of the 1/8 BPS
Bekenstein-Hawking black hole entropy starting with the formula
for the degeneracy of states in string theory, following
\cite{Maldacena:1999bp,Shih:2005qf,Pioline:2005vi,Sen:2009gy}. In
string theory various computations of the degeneracy of 1/8 BPS
states corresponding to a set of black holes charges $\cQ$ have
been performed. Charges are quantized and are often considered to
be `primitive', \ie such that their ``gcd'' (greatest common
divisor) equals 1. In the past, the goal of such counting formulae
was to derive the U-duality invariant 1/8 BPS black hole entropy
formula by means of the microscopic counting of states and provide
the quantum corrected formula for the degeneracy of states with
small charges, for which the classical black hole approximation
fails. A formula for the exact degeneracies of 1/8 BPS black holes
was originally derived in
\cite{Maldacena:1999bp,Shih:2005qf,Pioline:2005vi}, and is given
by \be d_{\cN =8}(\cI_4(\cQ))= \oint d\tau F(\tau) e^{-2\pi i
\tau\, \cI_4(\cQ)} \ee where the modular form \be F(\tau)=
{\theta_3(2\tau)\over \eta^6(4\tau)} \ee is written in terms of
the Jacobi theta function \be \theta_3(q) = \sum_n q^{n^2/2}
\label{eta} \ee with $q=e^{2\pi i \tau}$ and \be \eta(q) =
q^{1/24}\prod_n (1-q^n)\ee Dedekind's eta function.  In order to
determine the asymptotic behavior of $F(\tau)$ it is convenient to
perform a modular transformation \be \eta(-1/\tau) = (i\tau)^{1/2}
\eta(\tau) \qquad \theta_3(-1/\tau) = (i\tau)^{1/2} \theta_4(\tau)
\ee where\be \theta_4(q) = \sum_n (-)^n q^{n^2/2} \ee in this way
\be F(\tau) = {\theta(-1/2\tau) (4i\tau)^3 \over
(2i\tau)^{1/2}\eta^6(-1/4\tau) }\approx 2^{11/2} i^{5/2}
\tau^{5/2} e^{i\pi/8\tau} (1 + (6-2) e^{-i\pi/2\tau} + ...)\ee The
dots represent higher powers of $e^{-\pi i/2\tau}$ in the
expansion. In \cite{Sen:2009gy} this expression was derived for
large values of the quartic invariant and large values of the
charges. In such a case, one only considers the leading
exponential term and performs the integral over $\tau$ using a
saddle-point approximation. After taking into account the
determinant that comes from the integral around the saddle point
and evaluating the determinant (together with the integrand) at
the saddle point one finds \cite{Sen:2009gy}: \be d_{\cN
=8}(\cI_4(\cQ)) \sim (-1)^{\cI_4(\cQ)+1}  (\cI_4(\cQ))^{-2} e^{\pi
\sqrt{\cI_4(\cQ)} } \label{SSY}\ee Thus, for large black hole
horizon area $\cI_4(\cQ)$ and large charges $\cQ$ the quantum
mechanical counting of states in string theory agrees with the
Bekenstein-Hawking entropy of the black holes, \ie \be d_{\cN
=8}(\cI_4(\cQ)) \sim e^{S_{BH}(\cQ)+...} \label{BH} \ee where $
S_{BH} = {1\over 4} A_H = \pi \sqrt{|\cI_4|} $
\cite{Kallosh:1996uy} and $ \cI_4 = q^{abcd} \cQ_a \cQ_b \cQ_c
\cQ_d $ is the quartic invariant of $E_7$ and $\cQ_a$ is a
56-dimensional vector of electric and magnetic charges. The
expansion is valid for large charges and area of the horizon,
where the classical approximation is valid. It  may be viewed as a
limit when the occupation numbers of the quantum states are large
and the system is semi-classical. It is important, however, that
for small occupation numbers \ie small charges and vanishing area,
the above expansion it not valid.

To understand the properties of the degeneracy formula for states
with small charges and vanishing horizon area $\cI_4(\cQ)=0$ of
the corresponding ``black holes'',  we have to study a recently
developed version of eq. (\ref{SSY}) presented in
\cite{Sen:2009gy}, where the dependence on discrete
$E_{7(7)}(\mathbb{Z})$ invariants is introduced.

\
\section{Orbits and Discrete Invariants of $E_{7(7)}(\mathbb{Z})$}

Classical $\cN =8$ supergravity in $D=4$ enjoys a non-compact
$E_{7(7)}(\mathbb{R})$ symmetry. This continuous symmetry is
broken down to the discrete subgroup $E_{7(7)}(\mathbb{Z})$, when
the electric and magnetic charges of black hole solutions are
quantized \cite{Hull:1994ys}.

Meanwhile, the orbits of exceptional non-compact continuous
$E_{7(7)}(\mathbb{R})$ group were studied in
\cite{Ferrara:1997ci}, \cite{Ferrara:1997uz}.  These orbits are in
close parallel to the orbits of time-like, light-like and
space-like vectors in Minkowski space. The difference is that no
quadratic norm, analogous to $p^2= p_\mu p_\nu \eta^{\mu\nu}$ in
$SO(1,3)$,  is  available for $E_{7(7)}$, there is only a quartic
invariant $\cI_4(\cQ)$ depending on 56 charges $\cQ$.

The ``time-like'' orbit in $E_{7(7)}(\mathbb{R})$ corresponds to
$\cI_4>0$, the ``light-like'' to  $\cI_4=0$, and the
``space-like'' to  $\cI_4<0$. The ``time-like'' orbit of
$E_{7(7)}(\mathbb{R})$ with $\cI_4>0$ defines a set of regular
black holes with 1/8 unbroken supersymmetry. The corresponding
minimally short multiplet is generated by 28 fermionic zero modes
of the broken supersymmetry.  The ``space-like'' orbit of
$E_{7(7)}(\mathbb{R})$ with $\cI_4<0$ defines a set of regular
extremal non-BPS black holes. The corresponding long multiplet is
generated by the 32 fermionic zero modes of the broken
supersymmetry. Alternatively, it may be viewed as a bound state of
a set of two 1/2 BPS black holes with residual 1/4 unbroken
supersymmetry,  in such a way that the extra 4 fermion zero modes
can be used to form the bound state, as explained in
\cite{Sen:2008ta}.

The  light-like orbit with  $\cI_4=0$ presents three distinct
cases. The distinction between them is $E_{7(7)}(\mathbb{R})$
invariant. Despite the fact that there is only one available
invariant, the extra conditions  distinguishing among three types
of $\cI_4=0$ states are invariant: they are specified by the
vanishing of particular covariant tensors of
$E_{7(7)}(\mathbb{R})$. As we will show below, there is no
U-duality transformation which will mix the three cases with
$\cI_4=0$.

Here it is instructive to recall that bilinears of two {\bf 56}
decomposes according
 \be
  ({\bf 56} \otimes {\bf 56})_{\rm Symm} = {\bf 133} \oplus {\bf 1463}  \ee
  \be
  ({\bf 56} \otimes {\bf 56})_{\rm Anti} = {\bf 1} \oplus {\bf 1539}  \ee
where ${\bf 133}$ is the Adjoint.

The first case in the light-like (double critical) orbit  has 1/2
unbroken supersymmetry and the corresponding   ultra-short
multiplet is generated by the 16 fermionic zero modes of the
broken supersymmetry. 1/2 BPS-ness requires
 \be
\cP^{(ab)}_{\bf 133} {\partial \cI_4\over
\partial \cQ_a \partial \cQ_b} =0 \label{1/2} \ee
 which also implies $\cI_4=0$ and ${\partial \cI_4\over
\partial \cQ_a}=0$.  Note that the vanishing of the projection into the Adj,
 $\cP^{(ab)}_{\bf 133} {\partial \cI_4\over \partial  \cQ_q
 \partial \cQ_b}=0$, is an $E_{7(7)}(\mathbb{R})$ invariant requirement.

The second case in the light-like (critical) orbit has 1/4
unbroken supersymmetry  and the corresponding `very'  short
multiplet is generated by the 24 fermionic zero modes of the
broken supersymmetry. 1/4 BPS-ness requires
 \be {\partial \cI_4\over \partial \cQ_a} =0 \ ,
 \qquad \cP^{(ab)}_{\bf 133} {\partial \cI_4\over \partial \cQ_q \partial \cQ_b} \neq 0
\label{1/4} \ee
 which also implies $\cI_4=0$. Here again, the vanishing of
 ${\partial \cI_4\over \partial \cQ_a}$ is an $E_{7(7)}(\mathbb{R})$
 invariant requirement.

The third case in the light-like orbit has 1/8 unbroken
supersymmetry and the corresponding  minimally short multiplet is
generated by the 28 fermionic zero modes of the broken
supersymmetry. Light-like 1/8 BPS-ness requires that
 \be \cI_4=0 \ ,  \qquad {\partial \cI_4\over \partial \cQ_a}
 \neq 0 \ , \qquad \cP^{(ab)}_{\bf 133}{\partial \cI_4\over
 \partial \cQ_a \partial \cQ_b} \neq 0
 \ee

Besides $\cI_4$, the reduced discrete $E_{7(7)}(\mathbb{Z})$
symmetry allows for more arithmetic invariants  which define the
physical states in string theory \cite{Sen:2008sp}. A   general
discussion of discrete invariants based on Jordan algebras and
Freudenthal duals is presented in \cite{K,Borsten:2009zy}. Here we
will  use the fact that the tensors, derivatives of the quartic
invariant, defining distinct orbits, transform covariantly under
$E_{7(7)}(\mathbb{R})$.

In particular, for a (non-compact) discrete symmetry one can
introduce the notion of gcd (greatest common divisor) of a finite
set of not all zero integers, \ie the greatest integer that
divides them all. By definition a gcd is positive. {\it The
discrete $E_{7(7)}(\mathbb{Z})$ invariants are given by the gcd of
certain sets of numbers which correspond to covariant tensors of
$E_{7(7)}(\mathbb{R})$}. This is the reason why there is a clear
relation between the distinct light-like orbits of
$E_{7(7)}(\mathbb{R})$ and discrete $E_{7(7)}(\mathbb{Z})$
invariants.

One can define a number of  discrete U-duality invariants besides
the quartic invariant $\cI_4(\cQ)$. Introducing $\tilde\cQ_a = \de
\cI_4(\cQ)/\de\cQ_a$, that are cubic in $\cQ_a$, and following
\cite{K,Sen:2008sp,Borsten:2009zy}, one has \bea
&& a_1(\cQ) = {\rm gcd} \{\cQ_a\} \label{d1inv}\\
&& a_2(\cQ) = {\rm gcd} \{\cP^{(ab)}_{\bf 133}{\de^2
\cI_4(\cQ)\over \de\cQ_a\de\cQ_b}\}
\equiv \psi(\cQ) \label{d2'inv}\\
&& a_3(\cQ) = {\rm gcd} \{\tilde\cQ_a \} \label{d3inv}\\
&& a_4(\cQ) = \cI_4(\cQ) \label{d4inv}\\
&& a'_4(\cQ) = {\rm gcd} \{\cP^{(ab)}_{\bf 1539}\cQ_a \tilde\cQ_b
\}\equiv \chi(\cQ) \label{d4'inv}\eea where the subscript denote
the order in $\cQ_a$, for example $a_1$ is linear in $\cQ_a$,
$a_3$ is cubic in $\cQ_a$ etc. The set of discrete invariants
presented above contains all the ones which are relevant for our
analysis, namely we have a gcd of a set of integer charges, ${\rm
gcd} \{\cQ_a\}$, a gcd of a set of bilinears of these charges,
$\psi(\cQ) $. Next we have  a gcd of a set of charges cubic in the
original ones, $ {\rm gcd} \{\tilde\cQ_a \}$ and  a quartic
invariant of $E_{7(7)}(\mathbb{R})$, $\cI_4(\cQ)$, which is also a
quartic invariant of $E_{7(7)}(\mathbb{Z})$. Finally we have a gcd
of a set of charges quartic in $\cQ_a$, $\chi(\cQ)$.

The list of discrete invariants above may be incomplete, for
example other terms like $a'_2(\cQ) ={\rm gcd} \{\cP^{(ab)}_{\bf
1463}{\de^2 \cI_4(\cQ)\over \de\cQ_a\de\cQ_b}\} $ may be studied
in this respect. This is beyond the scope of our current work.

For the discrete $E_{7(7)}(\mathbb{Z})$ the quartic invariant is
quantized and takes the values \cite{K,Borsten:2009zy} \be
\cI_4(\cQ) \in \{0, 1\} \qquad \rm mod \, 4 \ee $a_1(\cQ) = {\rm
gcd} \{\cQ_a\}$ is obviously linear in the charges $Q_a$,
describing the state, while $a_3(\cQ) = {\rm gcd} \{\tilde\cQ_a
\}$ is cubic in the charges $Q_a$.

The states with non-vanishing quartic  invariants $\cI_4(\cQ)\neq
0$ which are 1/8 BPS or non-BPS may have all other discrete
invariants non-vanishing. This means that they can be classified
according to the values of these invariants. For example, one
finds that the states in \cite{Sen:2009gy} (corresponding to a
particular choice of charges $\cQ$) are described by $\psi(\cQ)=1$
and generic integer, in fact even, values of $\chi(\cQ)$.

A more interesting situation occurs if one tries to classify the
states with $\cI_4(\cQ)= 0$ in terms of discrete invariants of
$E_{7(7)}(\mathbb{Z})$. Differently from $a_4= \cI_4(\cQ)$, which
is invariant even under $E_{7(7)}(\mathbb{R})$ transformations,
all other invariants are gcd's of certain sets of numbers which
correspond to some covariant tensors of $E_{7(7)}(\mathbb{R})$.
Prior to taking their gcd, these numbers would not even be
invariant under $E_{7(7)}(\mathbb{Z})$. {\it But all} gcd {\it are
positive by definition}, they never vanish (it does not make sense
to divide a set of numbers by zero). Note that the only
$E_{7(7)}(\mathbb{Z})$ invariant which can take zero value is
$\cI_4(\cQ)$ since it is not a gcd but a specific quartic function
of integer charges $\cQ$. It follows that:

All {\it 1/2 BPS states with $\cI_4(\cQ)= 0$ and $\cP^{(ab)}_{\bf
133}{\de^2 \cI_4(\cQ)\over \de\cQ_a\de\cQ_b}=0$ are excluded from
the classification in terms of discrete {\rm gcd} invariants
(\ref{d2'inv}) and (\ref{d4'inv})}. The condition $\cP^{(ab)}_{\bf
133}{\de^2 \cI_4(\cQ)\over \de\cQ_a\de\cQ_b}=0$ contradicts the
positivity of $\psi(\cQ)$. For 1/2 BPS states only $a_2(\cQ)\neq
0$.

All {\it 1/4 BPS states with $\cI_4(\cQ)= 0$ and $\tilde \cQ_a=
{\partial \cI_4\over \partial \cQ_a} =0$, $\cP^{(ab)}_{\bf
133}{\de^2 \cI_4(\cQ)\over \de\cQ_a\de\cQ_b}\neq 0$ are excluded
from the classification in terms of discrete {\rm gcd} invariants
(\ref{d4'inv})}. The condition   $\tilde \cQ_a=0$ contradicts the
positivity of $\chi(\cQ)$. For 1/4 BPS states only $a_2(\cQ)\neq
0$ and $a'_2(\cQ)\neq 0$.

Not all {\it 1/8 BPS states with $\cI_4(\cQ)= 0$ and $\tilde
\cQ_a\neq 0$, $\cP^{(ab)}_{\bf 133}{\de^2 \cI_4(\cQ)\over
\de\cQ_a\de\cQ_b}\neq 0$ are excluded from the classification in
terms of discrete {\rm gcd}  invariants} since all  positive
discrete invariants are still compatible with  $\cI_4(\cQ)= 0$.
These states form a class of degenerate orbits, consistent with
the arithmetic classification of states. They are described in
details in Sec. 4.5 of \cite{K} where the integral version of
Freudenthal's construction is developed.

Let us stress here that our conclusion on the decoupling of 1/2
and 1/4 BPS states from the classification using discrete
$E_{7(7)}(\mathbb{Z})$ invariants is in a complete agreement with
the definition of the supersymmetric index $B_{14}$ which is
computed in \cite{Sen:2009gy}.  The corresponding index is
U-duality invariant, according to \cite{Sen:2009gy}.  The states
with 1/4 unbroken supersymmetry would contribute to a different
index, $B_{12}$,  and  the 1/2 BPS states would contribute to a
separate index $B_{8}$.

\section{ Sen's Formulae for the Degeneracy of $\cN =8$ BPS States }

An exact formula for the degeneracy of $\cN =8$ states in string
theory was derived in \cite{Sen:2009gy,Sen:2008sp,Sen:2008ta}. The
leading contribution for large areas and charges reproduces the
black hole entropy formula (\ref{BH}). The exponentially
subleading contribution depends on discrete $E_{7(7)}(Z)$
invariants specifying the state.

The formula for the degeneracy of states has been derived in
\cite{Sen:2009gy,Sen:2008sp,Sen:2008ta} in the framework of type
IIB string theory compactified on a  $T^6 = T^4\times S^1 \times
\tilde{S}^1$. The states are described by  a system of  D5/D3/D1
branes wrapping 4/2/0 cycles of a $T^4$ times $S^1$ or
$\tilde{S}^1$ . Alternatively, they can be described in type IIA
theory as states in the NS-NS sector.

In the context of string theory, it is useful to consider the
split of U-duality into S-duality and T-duality subgroups
$E_{7(7)}(\mathbb{Z}) \supset  SL(2, \mathbb{Z}) \times SO(6,6;
\mathbb{Z})$.  In the decomposition $E_{7(7)} \rightarrow
SO(6,6)\times SL(2)$ one has ${\bf 56} \rightarrow ({\bf 12}, {\bf
2}) + ({\bf 32}, {\bf 1})$ \ie $\cQ_a \rightarrow (P_i, Q^i ;
{S}^\alpha)$.

In  \cite{Sen:2009gy,Sen:2008sp,Sen:2008ta} only $(\mathbf{8},
\mathbf{2})$ out of the $(\mathbf{12}, \mathbf{2})$ charges $Q_i,
P^i$ are taken into account, whereas the 32 $S_\alpha$
(corresponding to the R-R sector in the `standard' T-duality basis
but not in the basis chosen in
\cite{Sen:2009gy,Sen:2008sp,Sen:2008ta}) are set to zero. As a
result, the string theory method of state counting
\cite{Sen:2009gy,Sen:2008sp,Sen:2008ta} has a manifest $SL(2,
\mathbb{Z}) \times SO(4,4; \mathbb{Z})$ subgroup of the U-duality
symmetry.

The discrete invariants in the U-duality subgroup are defined in
\cite{Sen:2009gy,Sen:2008sp,Sen:2008ta} as follows: \be \ell_1 =
gcd \{Q_i P_j - Q_j P_i\}  ,  \ell_2 = gcd \{Q^2/2, P^2/2, Q\cdot
P\} \ee The degeneracy formula is known
\cite{Sen:2009gy,Sen:2008sp,Sen:2008ta} for the subset of
primitive charge vectors with $a_1(Q)=1$ for which also \be
   \gcd(\ell_1, \ell_2)=1
\label{l}\ee
For primitive charge vectors satisfying
this condition the degeneracy is given by
\be
d_{1/8 BPS} ^{\cN =8} (Q, P) =
(-)^{I_4(Q,P) +1 } \sum_{s \in  \mathbb{Z}, s | \ell_1, \ell_2}  s
\hat{c} (\cI_4(Q,P)/s^2) \label {d} \ee

 Any generalization would
necessarily put the generators of $SO(6,6)$ (\ie $\{Q_i P_j - Q_j
P_i + {S} \gamma_{ij} {S}\}$ in the $({\bf 66},{\bf 1})$ ) and
those of $SL(2)$ (\ie $\{Q^2/2, P^2/2, Q\cdot P\}$ in the $({\bf
1},{\bf 3})$) on the same ground as the remaining $({\bf 32},{\bf
2})$ ones $\{ P_i \gamma^i {S} , Q_i \gamma^i {S} \}$ in
$E_{7(7)}/ SO(6,6)\times SL(2)$, so as to form the full Adj of
$E_{7(7)}$ and define \be \ell_E = {\rm gcd}  \{ P_{\bf 133}^{ab}
\cQ_{a} \cQ_{b} \}\ee

The U-duality invariant generalization of the constraint on
charges (\ref{l}) is the requirement that \be \psi(\cQ) = \ell_E =
{\rm gcd} \{\cP_{\bf 133}^{ab} \cQ_{a} \cQ_{b} \} =1\ee The
physical states in the counting formula in
\cite{Sen:2009gy,Sen:2008sp,Sen:2008ta} in addition to satisfying
the constraint that $\psi(\cQ)=1$, may differ in the value of the
discrete U-duality invariant \be \chi(\cQ) = \tilde\ell_E = {\rm
gcd} \{\cP_{\bf 1539}^{ab} \cQ_{a} \tilde\cQ_{b} \} \ee where
$\tilde\cQ_{b} = \de \cI_4/\de\cQ^b$ are the ${\bf 56}$ `dual'
charges, cubic in the fundamental ones, introduced before. For the
models considered in \cite{Sen:2009gy,Sen:2008sp,Sen:2008ta} the
discrete invariant $\chi(Q, P)$ is even, see for example, eq. (24)
in \cite{Sen:2008sp}. An additional requirement on the Sen's
derivation is that it should be possible to rotate the charges to
lie inside the ({\bf 12},{\bf 2}) subspace.

The formula for the degeneracy of states with $a_1(\cQ)=1$ ,
$a_2(\cQ) = \psi(\cQ)=1$ and even $\chi(\cQ)$ can be written as
follows \footnote{Note that $(-1)^{Q\cdot P}$ can be written as
$(-1)^{I_4(\cQ)}$ since $I_4(\cQ)$  is odd (even) when $Q\cdot P$
is odd (even) as explained in \cite{Sen:2009gy}.}
 \be d_{1/8 BPS} ^{\cN =8} (\cQ) =
(-)^{I_4(\cQ) +1 } \sum_{s \in  \mathbb{Z}, 2s | \chi(\cQ)} s
\hat{c} (\cI_4(\cQ)/s^2) \label {d} \ee where $\hat{c} (n)\approx
(-)^{n+1} \exp(\pi\sqrt{n})/n^2$ is related to the Fourier
coefficients in the expansion \be {\vartheta^2_1(z|\tau) \over
\eta^6(\tau) } = \sum_{k,l} \hat{c} (4k -l^2) e^{2\pi i (k\tau + l
z)} \ee As a result the degeneracy formula (\ref{d})  is
manifestly U-duality invariant.

\subsection{Degeneracy formula for 1/8 BPS states with $I_4(\cQ)\neq 0$ }

First we consider the left hand side of the degeneracy formula
(\ref{d}), only derived in \cite{Sen:2009gy} for  states with
$\psi(\cQ)=1$ and $I_4(\cQ)\neq 0$.

On the right hand side we find an expression of the form \bea
&&d(\cQ)|_{ I_4(\cQ)\neq 0}^{\cN=8} =
 \nonumber \\
 \nonumber \\
&& (-)^{I_4(\cQ) +1 } \sum_{s \in  \mathbb{Z}, 2s | \chi(\cQ)} s
\hat{c} (\cI_4(\cQ)/s^2) \label {d1} |_{ I_4(\cQ)\neq 0}
\label{regular}\eea It is manifestly U-duality  invariant since it
depends on the quartic invariant and a discrete invariant
$\chi(\cQ)$ of $E_{7(7)}(\mathbb{Z})$. For large $\cI_4$  the $s=1
$ term dominates and the Bekenstein-Hawking formula for the black
hole entropy which depends only on the quartic invariant of
$E_{7(7)}(\mathbb{Z})$ is reproduced, as shown in eq. (\ref{BH}).
When $\cI_4(\cQ)$ is not large, the formula (\ref{regular}) is
valid and the answer depends on both non-vanishing $\cI_4(\cQ)$
and $\chi(\cQ)$.

\subsection{Degeneracy formula for 1/8 BPS states with $I_4(\cQ)= 0$}

Here we consider the left hand side of the degeneracy formula
(\ref{d}) only for states with $I_4(\cQ)= 0$  and $\psi(\cQ)=1$
and some values of the discrete invariant  $\chi(\cQ)$. For 1/8
BPS states  with $I_4(\cQ)= 0$ we find a simple answer
 \be d(\cQ) |_{ I_4(\cQ)= 0}^{\cN=8} \; =
 \sum_{s \in  \mathbb{Z}, 2s | \chi(\cQ)} 2 s  = N[\chi(\cQ)] \label {singular}
 \label{I=0} \ee
where we used the fact that $\hat{c} (0)=-2$ that can be proven as
follows. The elliptic genus
\be \cE (z, \tau) = {\theta_1(z|\tau)^2\over
\eta^6(\tau)} = \sum_{k,l} \hat{c} (4k -l^2) e^{2\pi i (k\tau + l
z)}
\ee can be written more explicitly using the product
expansions \be \theta_1(z|\tau) = 2i q^{1/8} \sin(\pi z) \prod_n
(1-q^n)(1- y q^n)(1- y^{-1} q^n) \ee with $y=\exp(2\pi i z)$.
Setting $s_z = \sin(\pi z)$ one finds that \be \cE = - 4 s_z^2 \prod_n
\left(1 + {4 s_z^2 q^n\over (1-q^n)^2} \right)^2 \ee expanding in
powers of $q$ one finds \be \cE = - 4 s_z^2 (1 + 4 s_z^2 q + q^2 [12
s_z^2 + 16 s_z^4] + q^3 [ 32 s_z^2 + 64 s_z^4] + ...)\ee since
$\hat{c}(n)$ is only a function of $n=4k - \ell^2$ one can read
$\hat{c}(0)$ from the term with $q^0 \,  (k=0)$ and $\ell=0$. Using
$s_z^2 = (1-\cos(2\pi i z))/2$ one can thus confirm that $\hat{c}(0)
= -2$.

The degeneracy formula is U-duality invariant, it depends on the
value of the discrete invariant $\chi(\cQ)$, since the integer $
N[\chi(\cQ)]$ is given by the sum over $s$ which has to be taken
over only those integers which are factors of $\chi(\cQ)/2$. This
is indicated by the symbol $2s|\chi(\cQ)$ in the sum. Thus for
example if $\chi(\cQ)=2$, then only the $s=1$ term contributes. So
the sum, $ N[\chi(\cQ)]$, is always finite for a given charge
vector $\cQ$, but the actual result of the sum depends
non-trivially on the arithmetic properties of the charge vector,
\eg divisibility of $\chi(\cQ)$.

Note that had one first considered the semi-classical
approximation of large charges and occupation numbers of particles
and large area of the horizon of the regular black holes, the
answer would have been $\chi(\cQ)$ independent as shown in eq.
(\ref{BH}). Instead, for degenerate orbits, corresponding to
singular classical solutions, one should perform a fully quantum
mechanical analysis in order to derive the correct degeneracy of
such states. One should insert the vanishing value of the quartic
invariant directly in the exact formula for the degeneracy of
states (\ref{d}). This leads to the simple $E_{7(7)}(\mathbb{Z})$
invariant result shown in eq. (\ref{singular}).

\section{Counting 1/2 and 1/4 BPS states in perturbative Type II
superstrings}

The purpose of this section is to illustrate the fact, already
discussed above, that the U-duality modular invariant partition
function for 1/2 and 1/4 BPS states of $\cN =8$ string theory do
not mix with the one for 1/8 BPS states. Moreover in some case,
such perturbative BPS states can be interpreted as BH's
\cite{Dabholkar:1995nc}.

After toroidal compactification, the perturbative spectrum of Type
II superstrings contains massless, 1/2 BPS, 1/4 BPS, and long
multiplets but NO 1/8 BPS ones. The reason is that one needs
either R-R charges or K-K monopoles or NS5-branes (`H-monopoles')
to be added to perturbative states of type II string theory to get
1/8 BPS states.

\subsection{1/2 BPS in $\cN=8$}
In $D=10$, as a result of the GSO projection, the one-loop
partition function for Type IIB reads \cite{Kiritsis:2007zz} \be
\cZ = {|\theta_3^4 - \theta_4^4 - \theta_2^4 - \theta_1^4|^2 \over
4 |\eta^{12}|^2 }\ee where $ \eta(q) $   is a Dedekind's function
(\ref{eta}) and $\theta_\alpha$ with $\alpha =1,2,3,4$ are Jacobi
elliptic functions, see Appendix for the details. The partition
function vanishes thanks to Jacobi's identity \footnote{{\it
Aequatio identica satis abstrusa}}, which accounts for
supersymmetry. Modular invariance results after inclusion of the
bosonic and (super)ghost zero-modes, producing a factor ${V /
{Im\tau}^4}$, that nicely combines with the modular invariant
measure $d^2\tau/Im\tau^2$. The partition function can be
expressed in terms of the characters of the $SO(8)$ current
algebra (Little Group for massless states in $D=10$) at level
$\kappa = 1$ (denoted as $V_8$, $S_8$, $C_8$, $O_8$ for vector,
spinor, co-spinor and singlet conjugacy classes) \be \cZ =
{|\cQ|^2 \over |\eta^8|^2} \ee where $\cQ = V_8 - S_8 = ({\bf 8}_v
- {\bf 8}_s) q^{1/3} + massive$ is the super-character introduced
in \cite{Bianchi:1998vq}.

After toroidal compactification, the one-loop partition function
reads  \be \cZ = \sum_{{\bf m}, {\bf n}}
q^{{\ap\over 4} {\bf p}_{_L}^2} \bar{q}^{{\ap\over 4} {\bf
p}_{_R}^2} {|\theta_3^4 - \theta_4^4 - \theta_2^4 - \theta_1^4|^2
\over 4 |\eta^{12}|^2 }\ee where \be {\bf p}_{_{L/R}} =
(E^t)^{-1}({\bf m}+B {\bf n}) \pm {1\over \ap} E {\bf n} \ee with
$E^{\hat{i}}_i$ the 6-bein for the metric $G_{ij} =
\delta_{\hat{i}\hat{j}}E^{\hat{i}}_iE^{\hat{j}}_j$ and $B_{ij}$
the anti-symmetric tensor, \eg $G_{ij} = R^2\delta_{ij}$ and
$B_{ij}=0$ for a square torus. In this approximation one can
account only for the dependence on $36$ moduli fields in the NS-NS
sector, plus dilaton and axion. Perturbative computations are
insensitive to the 32 (pseudo)scalars in the R-R sector. At any
rate, degeneracy formulae should be independent of continuous
moduli and should be valid anywhere in the interior of moduli
space, since no jumping is possible in $\cN=8$.

The 256 massless states correspond to taking the ground states
(neither oscillators nor generalized momenta) for both Left and
Right movers \be \cZ_{m=0} = ({\bf 8}_v - {\bf 8}_s) ({\bf 8}_v -
{\bf 8}_s) = 128_B - 128_F\ee the minus sign accounts for the
different statistic of bosons and fermions \ie $\cZ$ is rather a
Witten index $\cI_W = tr(-)^F (q\bar{q})^H$ than a genuine
partition function.

1/2 BPS correspond to excitations of the ground states with only
generalized momenta \ie no oscillators \be \cZ_{{1/2 BPS}} = ({\bf
8}_v - {\bf 8}_s) ({\bf 8}_v - {\bf 8}_s) \sum_{{\bf m}, {\bf n}}
q^{{\ap\over 4} {\bf p}_{_L}^2} \bar{q}^{{\ap\over 4} {\bf
p}_{_R}^2}\ee and the level matching (\ie only states with the
same power of $q$ and $\bar{q}$ are physical ones) requires ${\bf
p}_{_L}^2= {\bf p}_{_R}^2$ \ie ${\bf m}\cdot {\bf n} = 0$. For
each internal direction, only KK momentum or winding but not both
are allowed. The degeneracy of such states is thus \be d_{1/2BPS
}^{\cN=8} ({\bf m},{\bf n}) =1\ee for given charges such that
${\bf m}\cdot {\bf n} =0$. Indeed for any choice of ${\bf m}$ and
${\bf n} =0$ such that ${\bf m}\cdot {\bf n} =0$ there is only one
KK multiplet. Actually the self-conjugate 1/2 BPS multiplet is
obtained combining the complex conjugate multiplets associated to
$({\bf m}, {\bf n})$ and $(-{\bf m}, -{\bf n})$ (very much as for
$W^\pm$ in SYM). By U-duality one expects that the same applies to
all 1/2 BPS states, which should include also wrapped branes and
KK monopoles.

The structure of 1/2 BPS multiplets is very simple. The spin of
the states runs from $S=0$ to $S=2$. The multiplicity of the
various spins are given by representations of $Sp(8)$ that rotates
the 8 real supercharges acting as raising (and as many as
lowering) operators in the multiplet. One indeed finds \be {\bf 1}
(S=2) + {\bf 8} (S=3/2) + {\bf 27} (S=1) + {\bf 48} (S=1/2) + {\bf
42} (S=0) \ee and it is easy to check that the total number of
states is 256, \ie 128 bosons and 128 fermions (up to the doubling
mentioned above in order to make the multiplet self-conjugate)

\subsection{1/4 BPS in $\cN=8$}

1/4 BPS states correspond to excitations of the Left (or Right)
mover ground states with generalized momenta AND Right (or Left)
mover oscillators.

The structure of 1/4 BPS multiplets is less simple than for 1/2
BPS ones. The spin of the states runs from $S=s_{min}$ to
$S=3+s_{min}$. The multiplicity of the various spins are given by
representations of $Sp(12)$ that rotates the 12 real supercharges
acting as raising (and as many as lowering) operators in the
multiplet. For the simplest case, $s_{min} = 0$, $s_{Max} =3$, one
indeed finds \be {\bf 1} (S=3) + {\bf 12} (S=5/2) + {\bf 65} (S=2)
+ {\bf 208} (S=3/2) + {\bf 429} (S=1) + {\bf 572} (S=3/2) + {\bf
429} (S=0) \ee and it is easy to check that the total number of
states is $4096=2^{12}$, \ie $2048 = 2^{11}$ bosons and $2048 =
2^{11}$ fermions (up to the doubling mentioned above in order to
make the multiplet self-conjugate)

Let us consider the case with Left movers in the ground state, see
also \cite{Kiritsis:2007zz}
 \be
\cZ^L_{{1/4 BPS}} = ({\bf 8}_v - {\bf 8}_s) {\theta_3^4 -
\theta_4^4 - \theta_2^4 - \theta_1^4 \over 2 \eta^{12}} \sum_{{\bf
m}, {\bf n}} q^{{\ap\over 4} {\bf p}_{_L}^2} \bar{q}^{{\ap\over 4}
{\bf p}_{_R}^2}\ee level matching (\ie only states with the same
power of $q$ and $\bar{q}$ are physical ones) in this case does
not requires ${\bf p}_{_L}^2= {\bf p}_{_R}^2$ but rather $\ap {\bf
p}_{_L}^2 = \ap {\bf p}_{_R}^2 + 4 \hat{N}_R $ \ie ${\bf m}\cdot
{\bf n} = \hat{N}_R $, where $\hat{N}_R\ge 1$ is the total
oscillator number with respect to the Right mover ground state. A
similar expression can be found for $\cZ^R_{{1/4 BPS}}$ after
exchanging Left and Right movers. Actually the self-conjugate 1/4
BPS multiplets (with R-mover oscillator modes) is obtained
combining the complex conjugate multiplets associated to $({\bf
m}, {\bf n})$ and $(-{\bf m}, -{\bf n})$ both with ${\bf m}\cdot
{\bf n} = \hat{N}_R \ge 1$. On the other hand the self-conjugate
1/4 BPS multiplets (with L-mover oscillator modes) is obtained
combining the complex conjugate multiplets associated to $({\bf
m}, -{\bf n})$ and $(-{\bf m}, {\bf n})$ both with $-{\bf m}\cdot
{\bf n} = \hat{N}_L \ge 1$. For each choice of $({\bf m}, {\bf
n})$ compatible with the level matching (\ie only states with the
same power of $q$ and $\bar{q}$ are physical ones)condition there
is only one state in the lattice sum. The degeneracy comes from
the oscillator modes.

At first look $\cZ^{L/R}_{{1/4 BPS}}$ vanish faster than the
manifest ${\bf 8}_v - {\bf 8}_s$ factor would suggest on account
of the extra broken supersymmetry (24 out of 32). In order to
by-pass the problem and count states instead of computing an
index, one can simply `twist' back (multiplying states by $(-)^F$)
the above expression in the Right (or Left) moving sector and
simply as well as correctly get \be \hat\cZ^L_{{1/4 BPS}} = ({\bf
8}_v + {\bf 8}_s) {\theta_3^4 - \theta_4^4 + \theta_2^4 +
\theta_1^4 \over 2 \eta^{12}} \sum_{{\bf m}, {\bf n}} q^{{\ap\over
4} {\bf p}_{_L}^2} \bar{q}^{{\ap\over 4} {\bf p}_{_R}^2}\ee One
would have expected to arrive at the same conclusion using the
appropriate helicity supertrace formula, but $B_{12}$ vanishes in
this case, due to extra fermionic zero-modes, see the
discussion\cite{Sen:2009md}.

Using Jacobi's {\it Aequatio} the `twisted' index, which is the
genuine partition function, can be rewritten as \be
\hat\cZ^L_{{1/4 BPS}} = ({\bf 8}_v + {\bf 8}_s) {\theta_2^4 \over
2 \eta^{12}} \sum_{{\bf m}, {\bf n}} q^{{\ap\over 4} {\bf
p}_{_L}^2} {\bar q}^{{\ap\over 4} {\bf p}_{_R}^2}\ee that allows
to extract the exact degeneracy of perturbative 1/4 BPS states in
Type II theories.  It is amusing to see that for the first excited
level corresponding to ${\bf m}\cdot{\bf n} =1$ one has indeed
4096 states as required for a 1/4 BPS multiplet with spin ranging
from $s_{min} = 0$, $s_{Max} =3$. The precise field content can be
obtained by decomposing the product \be ({\bf 8}_v - {\bf
8}_s)^L_0 \times [({\bf 8}_v - {\bf 8}_c)_1 \times ({\bf 8}_v -
{\bf 8}_s)_0]^R \quad ,\ee where the subscript $0$ stands for the
(L- and R-mover) ground states and $1$ for the R-mover
oscillators, into representations of $SO(2)\times SO(6)$ and then
lifting $SO(2)$ (Little group of {\it massless} particles in
$D=4$, \ie helicity) to $SO(3)$ (Little group of {\it massive}
particles in $D=4$ \ie spin) \eg for the highest spin state with
$S=3$ one has $3^+ + 2^+ + 1^+ + 0 + 1^- + 2^- + 3^- \rightarrow
(S=3)$ and similarly for lower spins.

Using the asymptotic growth of the degeneracy of oscillator states
at (large) level $\hat{N}$ one finds \be d_{1/4} ({\bf m}, {\bf
n}) \approx \exp (2\pi \sqrt{2{\bf m}\cdot{\bf n}}) \ee that would
suggest an entropy/area \be S^{1/4 BPS}_{BH} = 2\pi \sqrt{2 {\bf
m}\cdot{\bf n}} \ee This formula is not U-duality invariant since
the lowest Cartan invariant $\cI_4$ is quartic. This is no
surprise. In order to derive it we had to twist the partition
function (Witten index) by an operator that acts non-trivially on
R-R charges. Thus although any 1/4 BPS state with two charges is
U-duality equivalent to a perturbative state with momentum and
winding such that ${\bf m}\cdot{\bf n}\neq 0$, one cannot
immediately extend to non-perturbative states the above
perturbative degeneracy formula in this 1/4 BPS case.  See also
the discussion \cite{Sen:2009bm}.

In particular, if we blindly accept the above derivation of the
U-duality non-invariant entropy of the 1/4 BPS black holes, we
seem to have a contradiction with the fact that corrections, that
are quadratic in the charges and stretch the horizon, are absent
in D=4 $\cN=8$ supergravity. This is based on the fact that there
are no $R^2$ corrections in D=4 perturbative supergravity.
Equivalently, when IIB is reduced on $T^6$ and not on $K3\times
T^2$ there are no such corrections to the entropy (as different
from $\cN=4$ supergravity where such corrections are present and
are known \cite{Dabholkar:2004yr, Dabholkar:2004dq}).

\section{Resolution in D$>$4 of Singularities of  BPS states with $I_4(\cQ)= 0$  }

D=4 $\cN=8$ supergravity has  256 massless states which originate
from dimensional reduction of D=11 supergravity on $T^7$. To study
the non-perturbative states beyond the massless 256 we considered
extremal black holes in D=4 $\cN=8$ supergravity in
\cite{Bianchi:2009wj} and their masses.  They form five distinct
orbits of  $E_{7(7)}$ defined by the properties of the quartic
invariant $\cI_4(\cQ)$ and its derivatives with respect to the 56
charges $\cQ$. Namely, there are two orbits with $\cI_4>0$ and
$\cI_4<0$, the corresponding states have a Planck scale mass gap
and never become massless. There are three type of orbits with
$\cI_4(\cQ)=0$ describing 1/8, 1/4, 1/2 BPS states which may
become massless at the boundary\footnote{The boundary itself lies
at infinite distance from any interior point of the
$E_{7(7)}/SU(8)$ moduli space.} of the $E_{7(7)}/SU(8)$ moduli
space of D=4 $\cN=8$ supergravity. All such states correspond to
classical solutions with null singularities in D=4. We will study
these states by uplifting them to higher dimensions.

The uplifting from D=4 to D=5 may be viewed as a decomposition of
the U-duality group according to \cite{Ferrara:1997ci},
\cite{Ferrara:1997uz}, \cite{Ceresole:2007rq}
  \be E_{7(7)} \rightarrow E_{6(6)} \times O(1,1) \ee
\be \mathbf{56}\rightarrow \mathbf{27}_1 + \mathbf{1}_{-3} +
\mathbf{27}_{-1}' + \mathbf{1}_{3} \ee where in D=5 only $E_{6(6)}
$ is a symmetry.   The $ E_{6(6)}$ singlets are $p^0, q_0$ and the
pair of $\mathbf{27}$'s is $p^A, q_A$ with $A=1, ..., 27.$ In
order to explore the fate of the singularity of all extremal D=4
black holes, when they are uplifted to D=5, it is useful to
exploit the $E_{7(7)}$ decomposition of the D=4 black hole area of
the horizon in terms of the D=5 area of the horizon preserving the
$E_{6(6)}$ symmetry. This relation is known \cite{Ferrara:1997uz},
\cite{Pioline:2005vi}; \be \cI_4(\cQ)= - (p^0 q_0 + p^A q_A)^2 +
4\left [ p^0 \cI_3(q_A) -  q_0 \cI_3 (p^A) + {\partial
\cI_3(q_A)\over \partial q_A } {\partial \cI_3(p^A)\over \partial
p^A }\right ] \ee The extremal electric black holes and magnetic
strings  in D=5 have horizon area related to the cubic invariant
of the U-duality $E_{6(6)}$ symmetry
\cite{Ferrara:1996um},\cite{Ferrara:1997ci}
 \be
\cI_3(q_A)= d^{ABC} q_A q_B q_C \qquad
\cI_3(p^A)= d_{ABC} p^A p^B p^C
  \ee
The three distinct orbits in $E_{6(6)}$ describing the properties
of D=5 black holes were classified in \cite{Ferrara:1997uz},
\cite{Ferrara:1997ci}. All regular horizon extremal black holes in
D=5 have a non-vanishing cubic invariant \be \cI_3(q_A)\neq 0
\quad \Rightarrow 1/8 \rm BPS \ee The singular solutions are
either 1/4 or 1/2 BPS \be \cI_3(q_A) =  0 \quad  {\partial
\cI_3\over \partial q_A} \neq 0 \quad \Rightarrow 1/4 \rm BPS \ee
\be {\partial \cI_3\over \partial q_A}  = 0 \quad \Rightarrow 1/2
\rm BPS \ee in the second case, $\cI_3(q_A) =  0$ follows
immediately from the 1/2 BPS condition. Regular 1/8 BPS D=4 black
holes with $\cI_4\neq 0$ and singular 1/8 BPS D=4 black holes with
$\cI_4=0$ in D=5 are described by the 1/8 BPS regular black holes
with $ \cI_3(q_A)\neq 0$ or black strings with $\cI_3(p^A)\neq 0$.

{\it Thus the uplifting to D=5 removes the null singularity from
all 1/8 BPS extremal black holes in D=4 and makes them regular
extremal black holes in D=5}. Both regular and singular 1/8 BPS
black holes in D=4 become regular 1/8 BPS solutions in D=5. This
is in agreement with the fact that the $E_{6(6)}$ orbit with
$\cI_3 (q_A) \neq 0$ in D=5 splits into two distinct $E_{7(7)}$
orbits in D=4, one with $\cI_4 \neq 0$ and one with $\cI_4 = 0$.

A prototypical example in D=5 is a black hole depending on 27
electric charges with $\cI_3 \neq 0$ and $q_0, p^A=0$. In D=4 the
quartic invariant in such case is $\cI_4(\cQ)=  4 p^0 \cI_3(q_A)$.
It may be either zero, for $p^0=0$, or non-zero, for $p^0\neq 0$.
However, in D=5 both choices correspond to regular black holes
with $AdS_2\times S^3$ geometry near the horizon, whose area is
proportional to $\sqrt {\cI_3(q_A)}$.

The higher dimensional resolution of  black hole singularities was
studied in the past, particularly in \cite{Gibbons:1994vm}. It was
shown there that the singularity in the solutions of Einstein
theory with scalars in D=4 may be sometimes resolved when the
solution is viewed as a solution of Einstein theory in higher
dimensions. In particular, some extremal black holes with null
singularities become solutions with regular horizon in higher
dimensional space-times.

Thus we find that all singular 1/8 solutions,  which are massless
at the boundary of the moduli space in D=4, become regular black
holes with singularity covered by a horizon in D=5. This means
that one can cure all singular 1/8 BPS solutions by uplifting them
to D=5. In this way, however, one would loose the nice UV
properties of the perturbative D=4 $\cN=8$ supergravity. Indeed,
D=5 $\cN=8$ supergravity  is expected to be UV divergent, whereas
D=4 $\cN=8$ supergravity may turn out to be all-loop UV finite.

Now we consider singular 1/4 BPS solutions in D=4. In D=5 they are
still singular: the cubic $E_{6(6)}$ U-duality invariant vanishes,
$\cI_3(q_A) =0$.  To study the uplifting of $\cI_3=0$ black holes
to D=6 with U-duality group $O(5,5)$ we consider the following
decomposition \cite{Ferrara:1997ci}, \cite{Andrianopoli:2007kz}
$E_{6(6)}$ \be E_{6(6)} \rightarrow O(5,5) \times O(1,1) \ee The
$\mathbf{27}$ charges split as follows $\mathbf{27}\rightarrow
\mathbf{1}_4 + \mathbf{10}_{-2} + \mathbf{16}_1$, corresponding to
\be p^A= (p^z, p^r, p^\alpha ) \qquad q_A= (q_z, q_r, q_\alpha)
\qquad r= 1, ..., 10 \quad \alpha=1,..., 16. \ee Using this
splitting, the cubic invariant in D=5 can be related to quadratic
invariants in D=6 as follows
\cite{Ferrara:1997ci},\cite{Ferrara:2006yb} \be \cI_3(q_A)=
{1\over 2} \Bigl (q_z \; \cI_2(q_r) + q_r \;  q_\alpha
(\gamma^r)^{\alpha \beta} q_\beta \Bigr ) \ee where the quadratic
invariant  of the U-duality group $O(5,5)$ is \be \cI_2(q_r)=
\eta^{rs} q_r q_s \ee with $\eta^{rs}$ the $O(5,5)$ metric. In
this setting 1/4 BPS states are defined by the requirement that
the non-vanishing $\mathbf{10}$ vector is not null, or  a
non-vanishing $\mathbf{16}$ spinor $q_\alpha$ has a non-vanishing
vector bilinear $q_\alpha (\gamma^r)^{\alpha \beta}   q_\beta$
(\ie $q_\alpha$ is not a `pure' spinor). \be \cI_2(q_r) \neq 0
\quad {\rm or} \qquad q_\alpha (\gamma^r)^{\alpha \beta}
q_\beta\neq 0 \qquad \Rightarrow \qquad 1/4 BPS \ee The 1/4 BPS
solutions of D=6 maximal supergravity are dyonic strings. These
are regular solutions with $AdS_3\times S^3$ near horizon
geometry. The area of the horizon is proportional to $\sqrt
{\cI_2(q_r) }$. Thus, we see another example of a large class of
singular D=4 and D=5 black holes with $\cI_4=0$ and $ \cI_3=0$
which in D=6 become regular solutions. This class is described by
the vanishing singlet $q_z=0$, vanishing $\mathbf{16}$,
$q_\alpha=0$,  and non-vanishing non-null $\mathbf{10}$ vector.
These configurations with non-vanishing quadratic $O(5,5)$
invariant $\sqrt {\cI_2(q_r) }$ are regular in D=6.

{\it Thus the uplifting to D=6 removes the null singularity from a
class of 1/4 BPS extremal black holes in D=4 and makes them
regular dyonic strings in D=6}.
 The invariant $\cI_2(q_r)$ is very reminiscent of the 1/4 BPS
counting formula, since $ 4 \vec m\cdot \vec n= \ap
(p_L^2-p_R^2)$. In Type II strings in D=6, momenta and windings
span an $O(4,4)$ subspace of $O(5,5)$. The missing 8 charges are
accounted for either by wrapped branes (D1, D3 or D0, D2, D4) or
by uplift to M-theory, where there are 5 $p_L$ and 5 $p_R$ for
$T^5$.

We will not discuss here the intermediate cases of D=7,8,9 and
just note that in D=10 the well known 1/2 BPS D3 branes are
perfectly regular solutions. In D=11 the 1/2 BPS  M2 and M5 branes
are regular solutions, too.  But once again, we would like to
stress that D=10 and D=11 supergravities are not expected to be UV
finite and therefore the fact that non-perturbative solutions are
regular, may not be of particular significance since once quantum
effects are included one should resort to string or M theory for
consistent UV completions.

\section{On BPS \N=4 Yang-Mills   dyons and \N=8 supergravity black holes}

Although planar $\cN=4$ SYM and $\cN =8$ supergravity amplitudes
seem to secretly share the same structure, so much so that UV
divergences are actually absent to the order where computations
have been explicitly performed \cite{Bern:2009kf}, the structure
of their moduli space is rather different. While in $\cN =8$
supergravity giving VEV's to scalars does not break any symmetry,
the moduli space $E_{7(7)}/SU(8)$, being a symmetric space, is
homogeneous and any point is equivalent to the origin, in $\cN =4$
SYM only at the origin of the moduli space $R^{6r}/W_{r}$, with
$W_r$ the Weyl group of the rank $r$ gauge group, the theory
enjoys unbroken superconformal symmetry\footnote{Notice that at
this point, $\cN =4$ SYM has the same amount of supersymmetries as
$\cN=8$ supergravity, the extra (superconformal) charges are
precisely the ones needed to balance the supercharges in $\cN=8$
gauged supergravity in $D=5$ and to make the holographic
correspondence kinematically sensible.}. At any other point gauge
symmetry and superconformal symmetry are spontaneously broken and
an infinite tower of stable BPS dyons appear in the spectrum that
play a crucial role in the $SL(2,Z)$ e-m duality of \N=4 SYM
theory.

It is interesting to compare the role of non-perturbative states
in \N=8 supergravity and in \N=4 SYM theory at this point. \N=4
perturbative YM theory is UV finite, {\it per se}, when only the
16 massless states of the perturbative theory are included.
Non-perturbative monopoles and dyons are solutions of the
non-linear classical equations that essentially decouple from
perturbation theory. The proof of finiteness is based on the
properties of the vertices of the classical action where only 16
elementary massless states enter and there is no infinite tower of
light/massless dyons.

Although at the conformal point, infinitely many 1/2 BPS dyons
approach  the zero mass limit in \N=4 Yang-Mills theory, they
don't behave as elementary objects in the following sense. In the
conformal limit at weak coupling dyon masses go to zero, but their
size $L$ diverges even faster. Indeed, the Compton wavelength of
the dyon is related to its mass by $l_{Compt}\sim {1\over
M}={g\over v}$, while the classical size is determined by the
Compton wave-length of the massive W-bosons $L\sim{1\over gv}$.
One finds that ${l_{Compt}\over L} = g^2$ and, for very small
coupling $g^2\ll 1$, one has $l_{Compt}\ll L$.

As observed in \cite{Bianchi:2009wj}, the properties of
non-perturbative black hole solutions in \N=8 supergravity depend
crucially on the value of $\cI_4$. Regular BPS and non-BPS
extremal black holes $\cI_4\neq 0$ have a mass gap, while
solutions with $\cI_4\neq 0$ are singular and can become massless.
For regular extremal black holes, assuming that the  ADM mass
defines the Compton wavelength, $l_{Compt} \sim {1\over M_{ADM}}$
while the classical size is related to the area of the horizon,
$L\sim {1\over M_{H}}$, we may use the attractor properties of the
BPS black holes which suggests that $M_{H}\leq M_{ADM}$ and
$l_{Comt}\leq L$. Thus one can see that the classical scale  is
larger or equal to the Compton scale. For the consistency of the
classical interpretation,  one should still require that   $M_H\gg
M_{Pl}$. In string theory the degeneracy of the relevant states
with $I_4\neq 0$ are described by formula (\ref{regular}). For
light or massless singular solutions with $I_4=0$ there is no
reasonable classical interpretation in D=4. In M/string theory we
have found that the formulae for the degeneracy of states are
different from the one for regular black holes.

In summary, UV finiteness of \N=4 SYM is well established, while
UV finiteness of \N=8 is still conjectural. Despite differences
between specific properties of the non-perturbative states in the
two theories, one may still take a lesson for the UV properties of
\N=8 supergravity from \N=4 perturbative and non-perturbative \N=
4 SYM theory.  UV finiteness of the perturbative QFT relies only
on the 16 massless physical states of the CPT-invariant
supermultiplet. Although \N=4 SYM theory has an infinite tower of
BPS monopoles and dyons with exact mass formula, that become
massless in the conformal limit, they anyway decouple from the
perturbations theory\footnote{We thank M.~Porrati for suggesting
this analogy.}.

\section{Discussion}

In this paper we tried to relate M-theory on $T^7$  and Type II
superstring theory on $T^6$ to D=4 \N=8 supergravity. M-theory on
$T^7$  and Type II superstring theory on $T^6$ have, in addition
to \N=8 D=4 supergravity multiplet,  an infinite number  of
elementary massless states \cite{Green:2007zzb}, which are counted
in the degeneracy of states formulae. D=4 \N=8 supergravity has
256 massless states only and the perturbative QFT includes in the
Feynman graphs only these states.

The recent conjecture of all-loop UV finiteness of D=4 \N=8
perturbative supergravity is based on the extrapolations of the
recent 3- and 4-loop computations to higher loop order
\cite{Bern:2009kd}. Relating these computations to similar ones in
\cN=4 Yang-Mills \cite{Bern:1998ug, Bianchi:2008pu} lends support
to the prediction that the critical dimension where the onset of
UV divergences is given by $D_c= 4+{6\over L}$, where $L$ is the
number of loops. The all-loop UV finiteness of D=4 \N=8
perturbative supergravity is also supported by the light-cone
supergraph prediction under condition that the
$E_{7(7)}(\mathbb{R})$ be valid at the quantum level
\cite{Kallosh:2009db}.

The opinion was expressed that the all-loop UV finiteness of D=4
\N=8 perturbative supergravity may contradict U-duality invariance
of the degeneracy formulae for states in M/string theory if the
light non-perturbative states were to decouple from the theory.
Such states were studied in \cite{Bianchi:2009wj} and were shown
to correspond to light-like orbits of the U-duality group with
vanishing quartic invariant $\cI_4(\cQ)=0$ and singular horizon,
when viewed as black holes in D=4. They come in 3 types, 1/8, 1/4,
1/2 BPS solutions. {\it A priori} one would expect that different
orbits of $E_{7(7)}(\mathbb{R})$ and $E_{7(7)}(\mathbb{Z})$ should
not mix with each other. Still one would like to see that there
are no anomalies by explicitly  studying the known formulae for
the degeneracy of BPS states.

A technical tool which allowed us to perform an important part of
the relevant analysis is the relation which we found between the
orbits of $E_{7(7)}(\mathbb{R})$ and discrete invariants of
$E_{7(7)}(\mathbb{Z})$. Namely we have established that the states
which are described by discrete Sen's invariants $a_2(\cQ)=
\psi(\cQ)$ and $a_4'(\cQ)= \chi(\cQ)$ are only 1/8 BPS. The ones
which are 1/2 and 1/4 BPS are not included into a set of states
classified by discrete invariants. The reason for this is that the
set of integers whose greatest common divisor  would define the
discrete invariant, has to vanish for the state to be in the 1/2
and 1/4 BPS orbits.  However, the greatest common divisor is
defined for a set of not all vanishing integers and is positive.
Thus formulae for the degeneracy of 1/2 and 1/4 BPS states must be
separate from the 1/8 BPS ones.

We presented  formulae for the degeneracy of 1/2 and 1/4 BPS
states with $a_4(\cQ)= \cI_4(\cQ)= 0$. We pointed out a puzzle,
already raised by Sen and associated with the degeneracy for 1/4
BPS states, viewed as perturbative string states or as \N=4 D=4
supergravity solutions. The 1/4 BPS solutions are singular with
vanishing area of the horizon since this is not expected to be
stretched by \N=8 quantum corrections in D=4. Meanwhile, the
string degeneracy formula would suggest a stretching of the
horizon quadratic in the charges. The corresponding twisted index,
however, is not U-duality invariant.

We have also shown that the degeneracy formula of 1/8 BPS states
\cite{Sen:2009gy} splits into two separate formulae, a U-duality
invariant one for $\cI_4(\cQ)\neq 0$ related to regular black
holes entropy for large values of  $\cI_4(\cQ)$. It is given in
eq. (\ref{regular}). The other one is a relatively simple exact
U-duality invariant formula for the degeneracy of 1/ 8 BPS states
with $\cI_4(\cQ)=0$ that depends on discrete invariants of
$E_{7(7)}(\mathbb{Z})$. It is given by eq. (\ref{I=0}) and
suggests that the degeneracy of 1/8 BPS states with $a_4(\cQ)=
\cI_4(\cQ)=  0$, $a_1(\cQ)=1$, $a_2(\cQ)=\psi (\cQ)=1$  depends on
the discrete $E_{7(7)}(\mathbb{Z})$ invariant $a_4'=\chi(\cQ)$ as
follows: \be d(\cQ)|_{\cI_4(\cQ)=  0}= \sum_{s \in  \mathbb{Z}, 2s
| \chi(\cQ)} 2 s \ee Thus we have shown explicitly that all
singular states with $\cI_4(\cQ)=  0$ are decoupled from the
degeneracy formula for regular black holes states with
$\cI_4(\cQ)\neq 0$ shown in eq. (\ref{regular}). The counting
formulae for 1/8 BPS regular black holes and 1/8 BPS singular ones
are separately U-duality invariant.

We have also explained that all singular $\cI_4(\cQ)= 0$ solution of \N=8 D=4 supergravity resolve the singularities when uplifted to $D>4$. There is an interesting situation here since in $D>4$ the maximal supergravity is  not expected to be UV
finite in perturbation theory and therefore one should resort to string or M theory for
consistent UV completions.

We compared \N=8 perturbative and non-perturbative  supergravity  with  \N=4 perturbative and non-perturbative Yang-Mills theory. The \N=4 YM QFT based on 16 massless physical states of the CPT-invariant supermultiplet is UV finite. Meanwhile, the theory also has the infinite tower of the BPS monopoles and dyons, in the conformal limit they become massless, but they do not affect the \N=4 perturbation theory. So, there is a precedent of decoupling of the non-perturbative states from the perturbation theory.

Based on  the decoupling in the degeneracy of states formulae of the singular non-perturbative
$\cI_4(\cQ)=0$ light states of  D=4 \N=8
supergravity, we are lead to conclude that our  study of the non-perturbative sector of the theory does not reveal any
contradiction with the conjectured all-loop
finiteness of D=4 \N=8 perturbative supergravity based on 256 massless states of the CPT-invariant supermultiplet.

\section*{Acknowledgments}

We aknowledge  discussions and useful correspondence with T.~Banks
and A.~Strominger and their critical remarks on our analysis in
\cite{Bianchi:2009wj} and suggestion to look at formulae for the
degeneracy of states with and without including singular states.
We are particularly grateful to A.~Sen for explaining his work on
arithmetic of $\cN =8$ black holes to us. We are grateful to
L.~Borsten, M.~Dine, L.~Dixon, M.~Duff, D.~Harlow,  A.~Linde,
A.~Marrani, J.~F.~Morales, M.~Porrati, J.~H.~Schwarz, D.~Simic,
S.~Shenker, L.~Susskind, V.~S.~Varadarajan and E.~Witten for their interest in the
problem and for stimulating discussions.
 This work was partially supported by the ERC Advanced
Grant n.226455 {\it ``Superfields''} and by the Italian MIUR-PRIN
contract 2007-5ATT78 {\it ``Symmetries of the Universe and of the
Fundamental Interactions''}. The work of SF has been supported in
part by DOE grant DE-FG03-91ER40662, Task C. The work of RK  is
supported by the NSF grant 0756174.

\section*{Appendix A}

Elliptic functions admit both infinite product and series
expansions, that we collect in this Appendix and read \bea
&&\theta_3(z|\tau) = \sum_{n=-\infty}^{+\infty} q^{{1\over 2}n^2}
y^n
\\
&&\theta_4(z|\tau) = \sum_{n=-\infty}^{+\infty} (-)^n q^{{1\over
2}n^2} y^n
\\
&&\theta_2(z|\tau) = \sum_{n=-\infty}^{+\infty} q^{{1\over
2}(n+{1\over 2})^2} y^{n+{1\over 2}}
\\
&&\theta_1(z|\tau) = i \sum_{n=-\infty}^{+\infty} (-)^n q^{{1\over
2}(n+{1\over 2})^2} y^{n+{1\over 2}}\eea where $y=\exp(2\pi i z)$
and \bea &&\theta_1(z|\tau) = 2i q^{1\over 8} \sin(\pi z)
\prod_{n=1}^\infty (1-q^n) (1-y q^n) (1-y^{-1} q^n) \\
&&\theta_2(z|\tau) = 2 q^{1\over 8} \cos(\pi z) \prod_{n=1}^\infty
(1-q^n) (1+y q^n) (1+y^{-1} q^n)\\
&&\theta_3(z|\tau) = \prod_{n=1}^\infty
(1-q^n) (1+y q^{n+{1\over 2}}) (1+y^{-1} q^{n+{1\over 2}})\\
&&\theta_4(z|\tau) = \prod_{n=1}^\infty (1-q^n) (1-y q^{n+{1\over
2}}) (1-y^{-1} q^{n+{1\over 2}})\eea

Elliptic functions provide a representation of the $SL(2,Z)$
modular group \be \eta(\tau +1) = e^{i\pi/12} \eta(\tau) \qquad
\eta(-1/\tau ) = \sqrt{i\tau} \eta(\tau)\ee \be \theta_3(\tau +1)
= \theta_4(\tau) \qquad \theta_3(-1/\tau ) = \sqrt{i\tau}
\theta_4(\tau)\ee

\be \theta_4(\tau +1) = \theta_3(\tau) \qquad \theta_4(-1/\tau ) =
\sqrt{i\tau} \theta_2(\tau)\ee \be \theta_2(\tau +1) =
-\theta_2(\tau) \qquad \theta_2(-1/\tau ) = \sqrt{i\tau}
\theta_4(\tau)\ee

\

\section*{Appendix B}

In this Appendix, we would like to study the singular  states in
D=4 from the perspective of M/string theory. For this purpose it
may be useful first to identify the black hole charges $\cQ$ in
M-theory compactified on a seven-torus  $T^7$. Here we follow
\cite{Pioline:2005vi} and identify the 28+28 electric and magnetic
charges $\cQ= Q+P$ as M2, M5 branes, KK momenta (KKp) and KK
monopoles (KKm). In more detail, one can construct two  8x8
antisymmetric matrices, $I, J= 1, ..., 7$: \be Q= +\left(
  \begin{array}{cc}
    [M2]^{IJ} & [KKm]^I \\
   -[KKm]^I & 0
  \end{array}
\right) \qquad P= +\left(
  \begin{array}{cc}
    [M5]_{IJ} & [KKp]_I \\
   -[KKp]_I & 0
  \end{array}
\right) \ee The 28 electric charges of the D=4 black holes $Q$
correspond to 21 different M2 branes wrapped on various $[IJ]$
directions and of 7 KKM wrapped on all internal directions but
$I$. The 28 magnetic charges of the D=4 black holes $P$ correspond
to 21 different M5 branes wrapped on various  directions but
$[IJ]$ and of 7 KK momentum along the internal directions  $I$. In
terms of these M-theory objects the extremal black hole area of
the horizon is $S_{BH, 4D}= \pi \sqrt {\cI_4(P, Q)}$ where \be
\cI_4(\cQ) = \cI_4(P,Q) = -\Tr (OPQP)+{1\over 4} (\Tr QP)^2 -4
[{\rm Pf}(P) + {\rm Pf}(Q)] \ee If one would like to associate the
D=5 black holes with the states in M theory compactified on a six
torus $T^6$, one can now split one particular direction out of
seven $I$ so that $I= \{1, i\} , \; i= 2,..., 7$. Now the  the 56
charges $\cQ$ are split into 1+27+1+$\overline {27}$ charges \be
 q_0= [KKp]_1, \qquad q_A= \{[M2]^{ij}, \; [KKp]_i, \; [M5]_{i1} \}
\label{electric}\ee
\be
 p^0= [KKm]^1, \qquad p^A= \{[M2]^{i1}, \;[KKm]^i, \;[M5]_{ij} \}
\ee Here $q_A$ are the 27 electric charges of the D=5 extremal
black holes and $p^A$ are the $\overline {27}$ magnetic charges of
the D=5 black strings.


\begin{thebibliography}{10}

%\cite{Bianchi:2009wj}
\bibitem{Bianchi:2009wj}
  M.~Bianchi, S.~Ferrara and R.~Kallosh,
``Perturbative and Non-perturbative N =8 Supergravity,''
  arXiv:0910.3674 [hep-th].
  %%CITATION = ARXIV:0910.3674;%%

  %\cite{Bern:2009kd}
\bibitem{Bern:2009kd}
  Z.~Bern, J.~J.~Carrasco, L.~J.~Dixon, H.~Johansson and R.~Roiban,
 ``The Ultraviolet Behavior of N=8 Supergravity at Four Loops,''
  Phys.\ Rev.\ Lett.\  {\bf 103}, 081301 (2009)
  %[arXiv:0905.2326 [hep-th]].
  %%CITATION = PRLTA,103,081301;%%
  %\cite{Kallosh:2009jb}
%\bibitem{Kallosh:2009jb}
  R.~Kallosh,
 ``On UV Finiteness of the Four Loop N=8 Supergravity,''
  JHEP {\bf 0909}, 116 (2009)
 % [arXiv:0906.3495 [hep-th]].
  %%CITATION = JHEPA,0909,116;%%
 %\cite{Bern:2007hh}
%\bibitem{Bern:2007hh}
  Z.~Bern, J.~J.~Carrasco, L.~J.~Dixon, H.~Johansson, D.~A.~Kosower and R.~Roiban,
``Three-Loop Superfiniteness of N=8 Supergravity,''
  Phys.\ Rev.\ Lett.\  {\bf 98}, 161303 (2007)
 % [arXiv:hep-th/0702112].
  %%CITATION = PRLTA,98,161303;%%
%\cite{Green:2006yu}
%\bibitem{Green:2006yu}
  M.~B.~Green, J.~G.~Russo and P.~Vanhove,
 ``Ultraviolet properties of maximal supergravity,''
  Phys.\ Rev.\ Lett.\  {\bf 98} (2007) 131602
  [arXiv:hep-th/0611273].
  %%CITATION = PRLTA,98,131602;%%

%\cite{Kallosh:2009db}
\bibitem{Kallosh:2009db}
  R.~Kallosh,
  ``N=8 Supergravity on the Light Cone,''
   Phys.\ Rev.\  D {\bf 80}, 105022 (2009)
  arXiv:0903.4630 [hep-th].
  %%CITATION = ARXIV:0903.4630;%%

%\cite{Bern:1998ug}
\bibitem{Bern:1998ug}
  Z.~Bern, L.~J.~Dixon, D.~C.~Dunbar, M.~Perelstein and J.~S.~Rozowsky,
  %``On the relationship between Yang-Mills theory and gravity and its
  %implication for ultraviolet divergences,''
  Nucl.\ Phys.\  B {\bf 530} (1998) 401
  [arXiv:hep-th/9802162].
  %%CITATION = NUPHA,B530,401;%%

%\cite{Bianchi:2008pu}
\bibitem{Bianchi:2008pu}
  M.~Bianchi, H.~Elvang and D.~Z.~Freedman,
  %``Generating Tree Amplitudes in N=4 SYM and N = 8 SG,''
  JHEP {\bf 0809} (2008) 063
  [arXiv:0805.0757 [hep-th]].
  %%CITATION = JHEPA,0809,063;%%

      %\cite{Green:2007zzb}
\bibitem{Green:2007zzb}
  M.~B.~Green, H.~Ooguri and J.~H.~Schwarz,
``Nondecoupling Supergravity from the Superstring,''
  Phys.\ Rev.\ Lett.\  {\bf 99} (2007) 041601
  [arXiv:0704.0777 [hep-th]].
  %%CITATION = PRLTA,99,041601;%%

%\cite{Schnitzer:2007rn}
\bibitem{Schnitzer:2007rn}
  H.~J.~Schnitzer,
``Reggeization of N=8 Supergravity and N=4 Yang-Mills Theory II,''
  arXiv:0706.0917 [hep-th].
  %%CITATION = ARXIV:0706.0917;%%


%\cite{Maldacena:1999bp}
\bibitem{Maldacena:1999bp}
  J.~M.~Maldacena, G.~W.~Moore and A.~Strominger,
 ``Counting BPS black holes in toroidal type II string theory,''
  arXiv:hep-th/9903163.
  %%CITATION = HEP-TH/9903163;%%

%\cite{Shih:2005qf}
\bibitem{Shih:2005qf}
  D.~Shih, A.~Strominger and X.~Yin,
  ``Counting dyons in N = 8 string theory,''
  JHEP {\bf 0606}, 037 (2006)
  [arXiv:hep-th/0506151].
  %%CITATION = JHEPA,0606,037;%%
  %\cite{Shih:2005he}
%\bibitem{Shih:2005he}
  D.~Shih and X.~Yin,
``Exact Black Hole Degeneracies and the Topological String,''
  JHEP {\bf 0604}, 034 (2006)
  [arXiv:hep-th/0508174].
  %%CITATION = JHEPA,0604,034;%%

%\cite{Pioline:2005vi}
\bibitem{Pioline:2005vi}
  B.~Pioline,
``BPS black hole degeneracies and minimal automorphic representations,''
  JHEP {\bf 0508}, 071 (2005)
  [arXiv:hep-th/0506228].





%\cite{Sen:2009gy}
\bibitem{Sen:2009gy}
  A.~Sen,
 ``Arithmetic of N=8 Black Holes,''
  arXiv:0908.0039 [hep-th].
  %%CITATION = ARXIV:0908.0039;%%

  %\cite{Sen:2008sp}
\bibitem{Sen:2008sp}
  A.~Sen,
``U-duality Invariant Dyon Spectrum in type II on T6,''
  JHEP {\bf 0808}, 037 (2008)
  [arXiv:0804.0651 [hep-th]].
  %%CITATION = JHEPA,0808,037;%%


%\cite{Sen:2008ta}
\bibitem{Sen:2008ta}
  A.~Sen,
``N=8 Dyon Partition Function and Walls of Marginal Stability,''
  JHEP {\bf 0807}, 118 (2008)
  [arXiv:0803.1014 [hep-th]].
  %%CITATION = JHEPA,0807,118;%%

%\cite{Kallosh:1996uy}
\bibitem{Kallosh:1996uy}
  R.~Kallosh and B.~Kol,
 ``E(7) Symmetric Area of the Black Hole Horizon,''
  Phys.\ Rev.\  D {\bf 53}, 5344 (1996)
 % [arXiv:hep-th/9602014].
  %%CITATION = PHRVA,D53,5344;%%

%\cite{Hull:1994ys}
\bibitem{Hull:1994ys}
  C.~M.~Hull and P.~K.~Townsend,
``Unity of superstring dualities,''
  Nucl.\ Phys.\  B {\bf 438}, 109 (1995)
 [arXiv:hep-th/9410167].
  %%CITATION = NUPHA,B438,109;%%
%\cite{Witten:1995ex}
%\bibitem{Witten:1995ex}
  E.~Witten,
``String theory dynamics in various dimensions,''
  Nucl.\ Phys.\  B {\bf 443}, 85 (1995)
[arXiv:hep-th/9503124].
  %%CITATION = NUPHA,B443,85;%%

  %\cite{Ferrara:1997ci}
\bibitem{Ferrara:1997ci}
  S.~Ferrara and J.~M.~Maldacena,
``Branes, central charges and $U$-duality invariant BPS
conditions,''
  Class.\ Quant.\ Grav.\  {\bf 15} (1998) 749
  [arXiv:hep-th/9706097].
  %%CITATION = CQGRD,15,749;%%


 %\cite{Ferrara:1997uz}
\bibitem{Ferrara:1997uz}
  S.~Ferrara and M.~Gunaydin,
``Orbits of exceptional groups, duality and BPS states in string
theory,''
  Int.\ J.\ Mod.\ Phys.\  A {\bf 13} (1998) 2075
  [arXiv:hep-th/9708025].
  %%CITATION = IMPAE,A13,2075;%%
%\cite{Lu:1997bg}
%\bibitem{Lu:1997bg}
  H.~Lu, C.~N.~Pope and K.~S.~Stelle,
  %``Multiplet structures of BPS solitons,''
  Class.\ Quant.\ Grav.\  {\bf 15} (1998) 537
  [arXiv:hep-th/9708109].
  %%CITATION = CQGRD,15,537;%%







\bibitem{K} S. Krutelevich,  Jordan algebras, exceptional groups,
and Bhargava composition,
J. Algebra 314 (2007) no. 2, 924977,
arXiv:math/0411104.


%\cite{Borsten:2009zy}
\bibitem{Borsten:2009zy}
  L.~Borsten, D.~Dahanayake, M.~J.~Duff and W.~Rubens,
``Black holes admitting a Freudenthal dual,''
  Phys.\ Rev.\  D {\bf 80}, 026003 (2009)
  [arXiv:0903.5517 [hep-th]].
  %%CITATION = PHRVA,D80,026003;%%

%\cite{Dabholkar:1995nc}
\bibitem{Dabholkar:1995nc}
  A.~Dabholkar, J.~P.~Gauntlett, J.~A.~Harvey and D.~Waldram,
 ``Strings as Solitons \& Black Holes as Strings,''
  Nucl.\ Phys.\  B {\bf 474} (1996) 85
  [arXiv:hep-th/9511053].
  %%CITATION = NUPHA,B474,85;%%

%\cite{Kiritsis:2007zz}
\bibitem{Kiritsis:2007zz}
  E.~Kiritsis,
  ``String theory in a nutshell,''
%\href{http://www.slac.stanford.edu/spires/find/hep/www?irn=7300247}{SPIRES entry}
{\it  Princeton, USA: Univ. Pr. (2007) 588 p}

%\cite{Bianchi:1998vq}
\bibitem{Bianchi:1998vq}
  M.~Bianchi, E.~Gava, F.~Morales and K.~S.~Narain,
``D-strings in unconventional type I vacuum configurations,''
  Nucl.\ Phys.\  B {\bf 547}, 96 (1999)
  [arXiv:hep-th/9811013].
  %%CITATION = NUPHA,B547,96;%%
  %\cite{Bianchi:1990tb}
%\bibitem{Bianchi:1990tb}
  M.~Bianchi and A.~Sagnotti,
``Twist symmetry and open string Wilson lines,''
  Nucl.\ Phys.\  B {\bf 361} (1991) 519.
  %%CITATION = NUPHA,B361,519;%%
%\cite{Bianchi:1990yu}
%\bibitem{Bianchi:1990yu}
  M.~Bianchi and A.~Sagnotti,
``On the systematics of open string theories,''
  Phys.\ Lett.\  B {\bf 247} (1990) 517.
  %%CITATION = PHLTA,B247,517;%%

%\cite{Sen:2009md}
\bibitem{Sen:2009md}
  A.~Sen,
 ``A Twist in the Dyon Partition Function,''
  arXiv:0911.1563 [hep-th].
  %%CITATION = ARXIV:0911.1563;%%


%\cite{Sen:2009bm}
\bibitem{Sen:2009bm}
  A.~Sen,
  ``Two Charge System Revisited: Small Black Holes or Horizonless Solutions?,''
  arXiv:0908.3402 [hep-th].
  %%CITATION = ARXIV:0908.3402;%%




 %\cite{Dabholkar:2004yr}
\bibitem{Dabholkar:2004yr}
  A.~Dabholkar,
``Exact counting of black hole microstates,''
  Phys.\ Rev.\ Lett.\  {\bf 94} (2005) 241301
  [arXiv:hep-th/0409148].
  %%CITATION = PRLTA,94,241301;%%

 %\cite{Dabholkar:2004dq}
\bibitem{Dabholkar:2004dq}
  A.~Dabholkar, R.~Kallosh and A.~Maloney,
``A stringy cloak for a classical singularity,''
  JHEP {\bf 0412} (2004) 059
  [arXiv:hep-th/0410076].
  %%CITATION = JHEPA,0412,059;%%

%\cite{Ceresole:2007rq}
\bibitem{Ceresole:2007rq}
  A.~Ceresole, S.~Ferrara and A.~Marrani,
 ``4d/5d Correspondence for the Black Hole Potential and its Critical
  Points,''
  Class.\ Quant.\ Grav.\  {\bf 24}, 5651 (2007)
  [arXiv:0707.0964 [hep-th]].
  %%CITATION = CQGRD,24,5651;%%

  %\cite{Ferrara:1996um}
\bibitem{Ferrara:1996um}
  S.~Ferrara and R.~Kallosh,
  ``Universality of Supersymmetric Attractors,''
  Phys.\ Rev.\  D {\bf 54}, 1525 (1996)
  [arXiv:hep-th/9603090].
  %%CITATION = PHRVA,D54,1525;%%

%\cite{Gibbons:1994vm}
\bibitem{Gibbons:1994vm}
  G.~W.~Gibbons, G.~T.~Horowitz and P.~K.~Townsend,
``Higher Dimensional Resolution Of Dilatonic Black Hole Singularities,''
  Class.\ Quant.\ Grav.\  {\bf 12}, 297 (1995)
  [arXiv:hep-th/9410073].
  %%CITATION = CQGRD,12,297;%%


  %\cite{Andrianopoli:2007kz}
\bibitem{Andrianopoli:2007kz}
  L.~Andrianopoli, S.~Ferrara, A.~Marrani and M.~Trigiante,
``Non-BPS Attractors in 5d and 6d Extended Supergravity,''
  Nucl.\ Phys.\  B {\bf 795}, 428 (2008)
  [arXiv:0709.3488 [hep-th]].
  %%CITATION = NUPHA,B795,428;%%


%\cite{Ferrara:2006yb}
\bibitem{Ferrara:2006yb}
  S.~Ferrara, E.~G.~Gimon and R.~Kallosh,
``Magic supergravities, N = 8 and black hole composites,''
  Phys.\ Rev.\  D {\bf 74}, 125018 (2006)
  [arXiv:hep-th/0606211].
  %%CITATION = PHRVA,D74,125018;%%


%\cite{Bern:2009kf}
\bibitem{Bern:2009kf}
  Z.~Bern, J.~J.~M.~Carrasco and H.~Johansson,
 ``Progress on Ultraviolet Finiteness of Supergravity,''
  arXiv:0902.3765 [hep-th].
  %%CITATION = ARXIV:0902.3765;%%




\end{thebibliography}
\end{document}